\documentclass[fleqn,usenatbib]{mnras}
\usepackage{newtxtext,newtxmath}
\usepackage{xcolor}

\usepackage[T1]{fontenc}
\usepackage[super]{nth}

\DeclareRobustCommand{\VAN}[3]{#2}
\let\VANthebibliography\thebibliography
\def\thebibliography{\DeclareRobustCommand{\VAN}[3]{##3}\VANthebibliography}

\usepackage{graphicx}	
\usepackage{amsmath}	

\newcommand{\swift}{\emph{Swift}}
\newcommand{\rxte}{\emph{RXTE}}

\title[MOOSE II: New Excursion and Spin-Up Acceleration]{Monitoring observations of SMC X-1's excursions (MOOSE)-II: A new excursion accompanies spin-up acceleration}

\author[Chin-Ping Hu et al.]{
Chin-Ping Hu,$^{1}$\thanks{E-mail: cphu0821@gm.ncue.edu.tw}
Kristen C. Dage,$^{2,3}$ 
William I. Clarkson,$^{4}$ 
McKinley Brumback,$^{5}$ 
Philip A. Charles,$^{6}$
\newauthor
Daryl Haggard,$^{2,3}$
Ryan C. Hickox,$^{7}$
Tatehiro Mihara,$^{8}$
Arash Bahramian,$^{9}$
Rawan Karam,$^{2,3}$
\newauthor
Wasundara Athukoralalage,$^{10}$
Diego Altamirano,$^{6}$
Joey Neilsen, $^{11}$
Jamie Kennea$^{12}$
\\
$^{1}$Department of Physics, National Changhua University of Education, Changhua, 50007, Taiwan\\
$^{2}$Department of Physics, McGill University, 3600 University Street, Montr\'eal, QC H3A 2T8, Canada\\
$^{3}$Trottier Space Institute at McGill, 3550 University Street, Montr\'eal, QC H3A 2A7, Canada\\
$^{4}$Department of Natural Sciences, University of Michigan-Dearborn, 4901 Evergreen Rd. Dearborn, MI, 48128, USA \\
$^{5}$Department of Astronomy, University of Michigan, 1085 S. University Ave. Ann Arbor, MI 48109, USA\\
$^{6}$Physics \& Astronomy, University of Southampton, Southampton, Hampshire SO17 1BJ, UK\\
$^{7}$Department of Physics \& Astronomy, Dartmouth College, 6127 Wilder Laboratory, Hanover, NH 03755, USA\\
$^{8}$MAXI team, Institute of Physical and Chemical Research (RIKEN), 2-1, Hirosawa, Wako, Saitama 351-0198, Japan\\
$^{9}$International Centre for Radio Astronomy Research $-$ Curtin University, GPO Box U1987, Perth, WA 6845, Australia \\
$^{10}$Department  of  Physics  and  Astronomy,  Michigan  State  University,  East Lansing, MI 48824, USA\\
$^{11}$ Villanova University, Department of Physics, Villanova, PA 19085, USA\\
$^{12}$Department of Astronomy and Astrophysics, The Pennsylvania State University, University Park, PA 16802, USA
}

\date{Accepted XXX. Received YYY; in original form ZZZ}

\pubyear{2022}

\begin{document}
\label{firstpage}
\pagerange{\pageref{firstpage}--\pageref{lastpage}}
\maketitle

\begin{abstract}
SMC X-1 is a high-mass X-ray binary showing superorbital modulation with an unstable period. Previous monitoring shows three excursion events in 1996--1998, 2005--2007, and 2014--2016. The superorbital period drifts from $\gtrsim60$ days to $\lesssim40$ days and then evolves back during an excursion. Here we report a new excursion event of SMC X-1 in 2020--2021, indicating that the superorbital modulation has an unpredictable, chaotic nature. We trace the spin-period evolution and find that the spin-up rate accelerated one year before the onset of this new excursion, which suggests a possible inside-out process connecting the spin-up acceleration and the superorbital excursion. This results in a deviation of the spin period residual, similar to the behaviour of the first excursion in 1996--1998.  In further analysis of the pulse profile evolution, we find that the pulsed fraction shows a long-term evolution and may be connected to the superorbital excursion. These discoveries deepen the mystery of SMC X-1 because they cannot be solely interpreted by the warped disc model. Upcoming pointed observations and theoretical studies may improve our understanding of the detailed accretion mechanisms taking place.
\end{abstract}




\section{Introduction}
SMC X-1 is a high-mass X-ray binary (HXMB) consisting of an accreting neutron star and a supergiant companion \citep{Reynolds1993, vanderMeer2007}. Its pulsation period is $0.7$ s and the source has steadily spun up since discovery, implying a Roche-lobe overflow stream-fed accretion mode \citep{Lucke1976, Inam2010, HuMS2019}. The orbital period of SMC X-1 is $3.89$ days and decays with a rate of $\dot{P}_{\rm{orb}}\approx3.8\times10^{-8}$ \citep{Wojdowski1998, Falanga2015, HuMS2019}. A quasi-periodic superorbital modulation of this system has long been observed in the X-ray band \citep[e.g.][]{Gruber1984, ClarksonCC2003a}. 

Radiation-driven warping is expected to be significant in stream-fed X-ray binaries with high central accretion luminosity and a physically large accretion disc \citep{Pringle1996, OgilvieD2001}. The disc of such a system is unstable to warping driven by the interception and re-radiation of accretion luminosity from the compact object and inner disc.
Our line of sight to the neutron star is then subject to obscuration by the warped region, leading to superorbital variation as the disc precesses. A few sources, such as Her X-1, LMC X-4, and MAXI J1820+070, are believed to have similar structures \citep{ClarksonCC2003b, ThomasCB2022}. This mechanism has been reproduced in hydrodynamical simulations of X-ray binaries \citep{FoulkesHM2010}. The stability analysis of \citet{OgilvieD2001} suggests that the character and precessional behaviour of a radiation-driven disc warp can be parameterized by the binary separation and mass ratio, leading to precession that could be stable, quasi-periodic, or aperiodic. A larger binary separation results in a more complicated behaviour, and SMC X-1 is in the regime of marginal instability of the warp \citep[e.g.][]{Charles2008}. 

Consistent with this prediction, the superorbital X-ray variability of SMC X-1 is known to not be strictly periodic. Variability in the timescale of its superorbital X-ray modulation was shown early in the lifetime of the RXTE mission \citep{Wojdowski1998}. The superorbital modulation of SMC X-1 is punctuated by superorbital excursion events (hereafter {\it superorbital excursions)}, where the superorbital period shortens from $\gtrsim60$ days to $\lesssim40$ days and then evolves back on a time scale of 2--3 years. 
Excursions have occurred at least three times in the intervals 1996--1998 (\nth{1}), 2005--2007 (\nth{2}), and 2014--2016 (\nth{3}) \citep{Trowbridge2007, Hu2011, DageCC2018, HuMS2019}.

SMC X-1 is a particularly valuable source because it shows {\it both} variations in superorbital modulation and in its pulse period, and thus the behaviour of the accretion engine and of its large, complicated accretion disc can be tracked together. For example, the spin-up rate increased during the \nth{1} excursion, but similar behaviour was not observed in the \nth{3} excursion \citep{Inam2010, DageCC2018, HuMS2019}. 

Recent studies on pulsating ultraluminous X-ray sources (PULXs), accreting pulsars with extremely high luminosities up to $\sim 100$~times the Eddington Luminosity, suggest that a significant fraction of PULXs exhibit superorbital modulations \citep{BachettiHW2014, KongHL2016, WaltonFB2016, HuLK2017, BrightmanEF2020}. Their superorbital modulations are argued to trace intrinsic changes in the mass accretion rate, and the propeller effect could be observed in these PULXs if their magnetospheric radii are similar to the co-rotation radius \citep{TsygankovMS2016, VasilopoulosHT2021}. Other models, such as those involving the precession of the disc or conical wind, are also suggested \citep{DauserMW2017, KingL2019}. Owing to its high peak luminosity (up to a few times Eddington), strong observed changes in the pulsed fraction, and possible intrinsic mass accretion rate variability, SMC X-1 could be a useful local analogue to PULXs \citep{PikeHB2019}. Investigating the superorbital modulation of SMC X-1 in detail is therefore not only a key to understanding the accretion physics but also may hint at  the nature of PULXs. 

All these special properties make SMC X-1 a unique source to study the instability of the accretion disc. Our collaboration has embarked on an intensive monitoring campaign to track the spectral-temporal variability of SMC X-1 with orbital and superorbital phase resolution, during and outside superorbital excursion. This campaign - {\bf M}onitoring {\bf O}bservations {\bf O}f {\bf S}MC X-1's {\bf E}xcursions (MOOSE) was introduced in \citet{DageBN2022}.

In this paper, the second in a planned series from the MOOSE campaign, we present the detection of a new (\nth{4}) excursion event $\sim1800$ days after the \nth{3} excursion, and explore its connection to the spin behaviour of SMC X-1. We describe the data reduction in section \ref{data}. The analysis results, including the time-frequency analysis of the superorbital modulation period, the spin period variability, and the change in pulse profiles, are described in section \ref{result}. We discuss the implications of these results in Section \ref{discussion} and summarize our work in section \ref{summary}.

\section{Data Processing}\label{data}

\subsection{Swift BAT}\label{bat_section}
The Burst Alert Telescope (BAT) onboard the Neil Gehrels Swift Observatory (hereafter \swift) has a large collecting area (5200\,cm$^2$). The \swift\ BAT hard X-ray transient monitoring program has recorded the X-ray count rate of known sources in the 15--50\,keV band since 2004, which makes it an ideal instrument to study the superorbital modulation \citep{BarthelmyBC2005, KrimmHC2013}. We use the one-orbit light curve (with a resolution of $96$ minutes), and eliminate data points that have uncertainties 3$\sigma$ higher than the mean uncertainty. We then remove the eclipse according to the orbital ephemeris presented in \citet{HuMS2019}, and re-bin the light curve with a one-day resolution. 

\subsection{RXTE ASM}
The All Sky Monitor (ASM) onboard the Rossi X-ray Timing Explorer (\rxte) consists of three proportional counter arrays with a collecting area of 90\,cm$^2$ \citep{LevineBC1996}. \rxte\ ended its mission in 2012, so we simply adopt data collected in \citet{HuMS2019}, which applied the same selection criteria as that described in Section \ref{bat_section} to investigate the superorbital modulation before MJD 55400. 

\subsection{MAXI GSC}
The Monitor of all-sky X-ray image (MAXI), a Japanese Experimental Module of the International Space Station (ISS), can monitor the entire sky in both X-ray (0.5--12 keV) and hard X-ray (2--30 keV) bands \citep{MatsuokaKU2009}. We search for pulsation using hard X-ray data collected with the Gas Slit Camera (GSC) that has a large collecting area of 5350\,cm$^2$ and a high time resolution of 50\,$\mu$s \citep{MiharaNS2011}. Among 12 proportional counters, we only use GSC IDs of 0, 1, 2, 4, 5, and 7 because the others are not in operation \citep{SugizakiMS2011}. After 2020 August, camera 0 is also closed due to gas leakage \citep{MihTN2022}. We extract X-ray photons in the 2--20\,keV band using \texttt{mxextract} from the MAXI database. The photon arrival times are corrected to the barycentre of the solar system according to the ephemeris DE-200. The source photons are extracted from a 1$^{\circ}$ radius circle centred on SMC X-1 that guarantees $\gtrsim90$\% of source photons are selected \citep{MiharaNS2011}. 

\begin{table*}
 \caption{Summary of superorbital excursions of SMC X-1}
 \label{tab:excursions}
 \begin{tabular}{lcccc}
 \hline
 Excursions & \nth{1} & \nth{2} & \nth{3} & \nth{4}\\
 \hline
 Onset MJD & 50400 & 53800 & 56850 & 58700\\
 Duration (days) & 650 & 400 & 600 & 850\\
 Interval$^a$ (days) & $\sim1300$--$1700$(?) & 3400 & 3050 & 1850\\
 Spin-up acceleration & Yes & $\ldots$ & No & Yes\\
 Low-state count Rate & Increase & Increase & Increase & Increase\\
 High-State count Rate & Stable & Stable & Stable & Stable\\
 Pulsed Fraction & $\ldots$ & $\ldots$ & Increase & Increase\\
 \hline
 \multicolumn{5}{l}{\small $^a$ Time interval between the onset of this excursion and that of the previous excursion.}\\
 \end{tabular}
\end{table*}

\section{Analysis and Results}\label{result}
\subsection{Superorbital Excursion}\label{ss:method:excursion}
The most recent analysis of the superorbital modulation of SMC X-1 was carried out by \citet{DageCC2018} and \citet{HuMS2019}. Here we extend the monitoring for an additional four years until 2022 August (MJD 59810). To trace the superorbital evolution, we adopt the stacked Hilbert-Huang transform (HHT) proposed in \citet{HuLP2022}. The HHT, a novel and continually-developed time-frequency analysis technique \citep{Huang1998, Wu2009}, has been successfully applied to the all-sky monitoring data of SMC X-1 \citep[see, e.g.,][]{Hu2011, HuMS2019}. 
The HHT algorithm is sensitive to data gaps. To address this issue, we use piecewise cubic Hermite interpolation to fill in gaps in the data and assume a Gaussian uncertainty with $\sigma$ equal to the mean value of other data points \citep{Kahaner1989}. The stacked HHT, which has been used to characterise the properties of gravitational wave signals, provides a good balance between maximising the resolution in both the time and frequency domains and eliminating possible spurious frequency modulation caused by noise and mode-splitting \citep[see][for details]{HuLP2022}. 

\begin{figure}
 \includegraphics[width=\columnwidth]{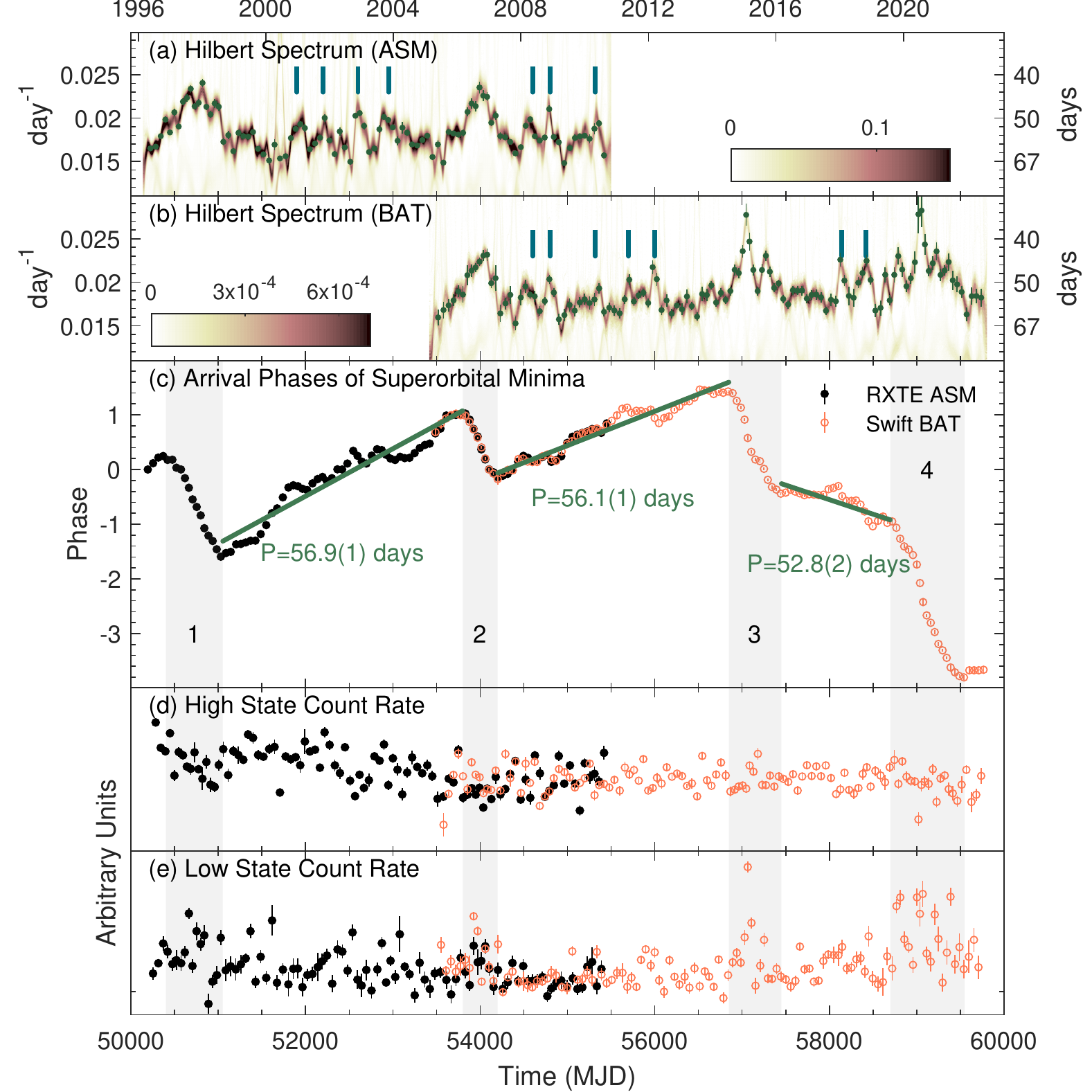}
 \caption{Stacked HHT spectra obtained with RXTE ASM (a), and Swift BAT (b), the phase evolution of the superorbital modulation (c), corresponding high-state (d) and low-state (e) count rates of SMC X-1. Gray shaded areas in panels (c)--(e) denote the time intervals of four excursion events (numbered 1--4). Colours in Hilbert spectra denote the Hilbert amplitude. Blue vertical lines in panels (a) and (b) indicate possible mini excursions. Green data points in panels (a) and (b) are the frequency of each superorbital cycle derived from the cycle length between two consecutive minima in panel (c). Green lines in panel (c) are best-fit linear models of three non-excursion epochs, where corresponding periods are labelled. The slight differences in the averaged periods between epochs are much smaller than the period change between superorbital cycles, and therefore cannot be seen in the Hilbert spectra. The high- and low-state count rates are scaled to roughly the same level for display purposes.}
 \label{fig:superorbital_shht}
\end{figure}

In brief, we perform $10^4$ Monte Carlo simulations. In each simulation, we create a light curve according to the observed count rates and implement white noise with standard deviations equal to the uncertainty. Then, we use the complementary ensemble empirical mode decomposition (EEMD) to decompose the light curves into a finite number of intrinsic mode functions \citep{YehSH2010}. This fast algorithm has been included in a Matlab package developed by the Research Center for Adaptive Data Analysis at National Central University \citep{WangYY2014}. Finally, we obtain $10^4$ time-frequency maps and stack them together. The resulting time-frequency maps obtained with both ASM and BAT data are shown in Figure \ref{fig:superorbital_shht} (a).

\begin{figure}
 \includegraphics[width=\columnwidth]{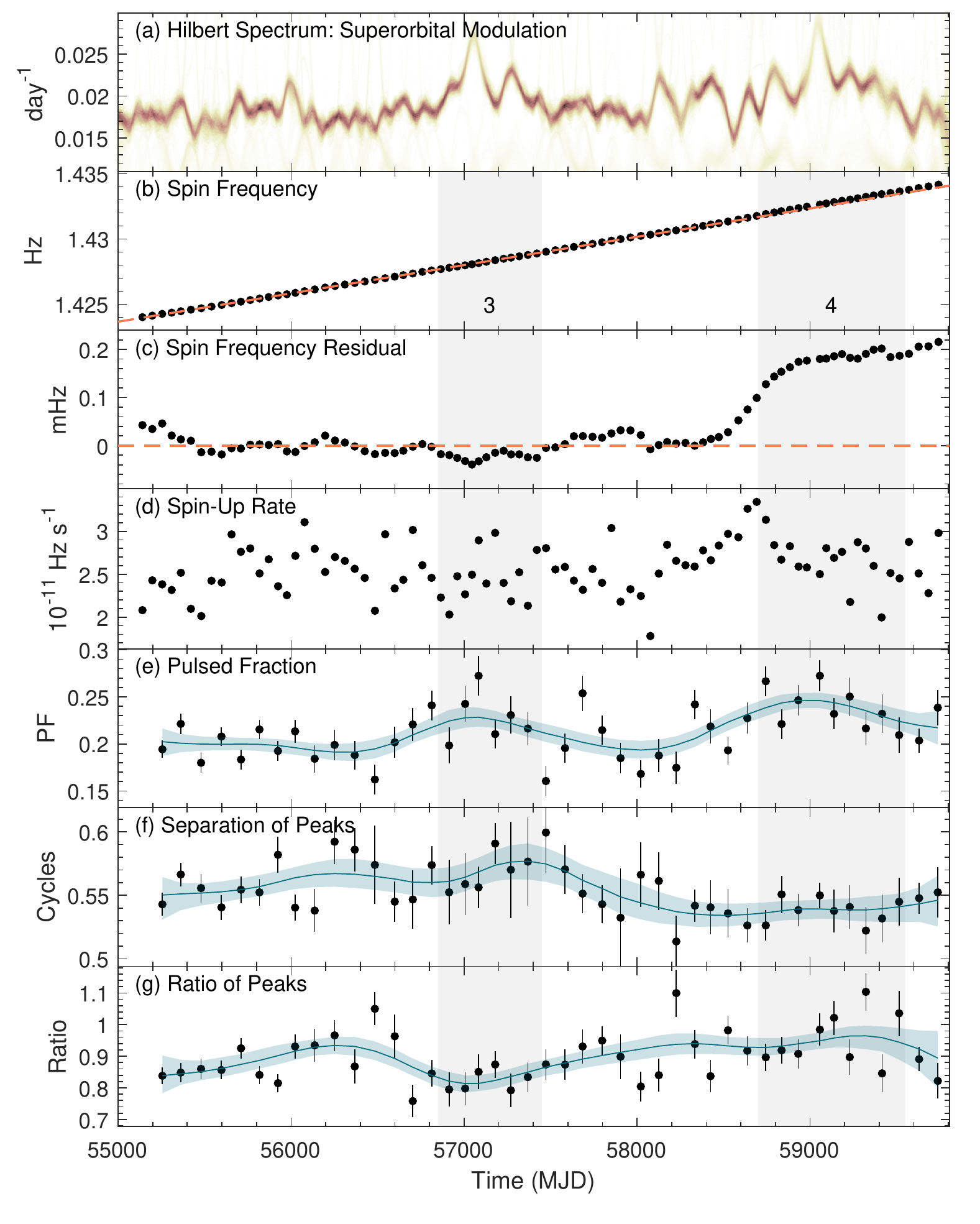}
 \caption{Evolution of the spin frequency and the pulse profile parameters of SMC X-1 obtained with MAXI. The stacked HHT spectrum obtained with Swift BAT is shown in panel (a) for reference. Panel (b) shows the spin period in each superorbital high state, where the orange dashed line denotes the best-fit linear model obtained using data before MJD 5,484. The residual after subtracting the linear trend and the spin-up rate are shown in panels (c) and (d), respectively. The evolution of the pulsed fraction (e), phase separation of two peaks (f), and the ratio of the two peaks (g) are shown in the following panels. Gray shaded area denotes the epochs of the \nth{3} and \nth{4} excursions. Blue curves are obtained from EEMD band-pass filtered data, where the light blue areas correspond to 1-$\sigma$ confidence intervals.}
 \label{fig:superorbital_spin}
\end{figure}

In addition, we trace the arrival times of the superorbital minima, which are obtained from the EEMD band-pass filtered light curve \citep[see][]{HuMS2019}. Assuming a folding period of 54.3 days, we performed a phase-coherent analysis on the arrival phase of the superorbital minima, which is shown in Figure \ref{fig:superorbital_shht} (c). We calculate cycle lengths between two consecutive minima and show the corresponding frequency in Figure \ref{fig:superorbital_shht}(a), suggesting that the result of the phase-coherent analysis is fully consistent with that of the HHT. From both the time-frequency map and the phase evolution, three known excursion events (\nth{1}--\nth{3}) are clearly seen. A new \nth{4} excursion occurred in 2020--2021 and is found in this analysis. The starting time of each excursion, which is defined as the starting point of phase drop, is listed in Table \ref{tab:excursions}. Time intervals between the onsets of the first three excursions are 3400 and 3050 days. However, the time interval between the onset of the \nth{3} and \nth{4} excursions is 1850 days, roughly half of the recurrent time scale observed in previous events. The baseline of the superorbital period in non-excursion epochs before the \nth{4} excursion seems to increase with time (Figure \ref{fig:superorbital_shht}). These suggest the unpredictable nature of the excursion of SMC X-1, and hint that SMC X-1 is entering a different mode of disc precession. 

Finally, we calculate the count rates in the superorbital high and low states \citep{HuMS2019, DageBN2022}. Since the energy ranges between ASM and BAT are quite different, we scaled and shifted the light curves to compare the relative variability, as shown in Figure \ref{fig:superorbital_shht} (d).  As described in \cite{HuMS2019}, the low-state count rate increased in the \nth{1}--\nth{3} excursions, but the high-state count rate remained stable. We find that the behaviour of high- and low-state count rates in the \nth{4} excursion agrees with previous events. The lack of change in the  high-state count rate suggests a stable mass accretion rate. On the other hand, the low-state count rate significantly increases during excursions, implying a possible change in the disc configuration.

\subsection{Spin Period Evolution}
Following the algorithm in \citet{HuMS2019}, we employ a two-dimensional $Z_2^2$ test \citep{Buccheri1983} to search for spin frequency ($\nu$) -- spin-up rate ($\dot{\nu}$) pairs within each superorbital high state and eliminate the orbital Doppler effects using MAXI GSC data. For each segment of data, we search for possible combinations of $\nu$--$\dot{\nu}$ pairs near the predicted value based on the result of the previous superorbital high state with an oversampling factor of 10. Then we use a finer resolution with an oversampling factor of 1000 to determine the peak location in the candidate detection. The determined $\nu$ and $\dot{\nu}$ are plotted in Figure \ref{fig:superorbital_spin} (b) and \ref{fig:superorbital_spin} (d). The uncertainties of $\nu$ and $\dot{\nu}$ are conservatively estimated from the width of the peak in the $Z_2^2$ spectrum. 

We see a monotonically increasing trend of the spin frequency. We display the local linear trend of the spin period evolution derived before 2019 (MJD 58484) in Figure \ref{fig:superorbital_spin} (b), and then obtain the residual in Figure \ref{fig:superorbital_spin} (c) to make a fair comparison with the evolution in \citet{HuMS2019}. The spin-up rate varies between neighbouring cycles in the range of 2 and $3\times10^{-11}$ Hz s$^{-1}$ through the entire MAXI observation. Interestingly, a significant spin-up rate increment, from 2.5 to $3.4\times10^{-11}$ Hz s$^{-1}$, is seen in MJD 58400 -- MJD 58700, just before the onset of the \nth{4} excursion. Then the source takes another 200 days to decrease back. This makes the spin frequency deviate rapidly from the linear trend as shown in Figure \ref{fig:superorbital_spin} (c). This phenomenon is not observed in the \nth{3} excursion but is similar to the offset between the spin period measurements and the linear trend observed in the \nth{1} excursion \citep{Inam2010, DageCC2018}.

\begin{figure}
 \includegraphics[width=\columnwidth]{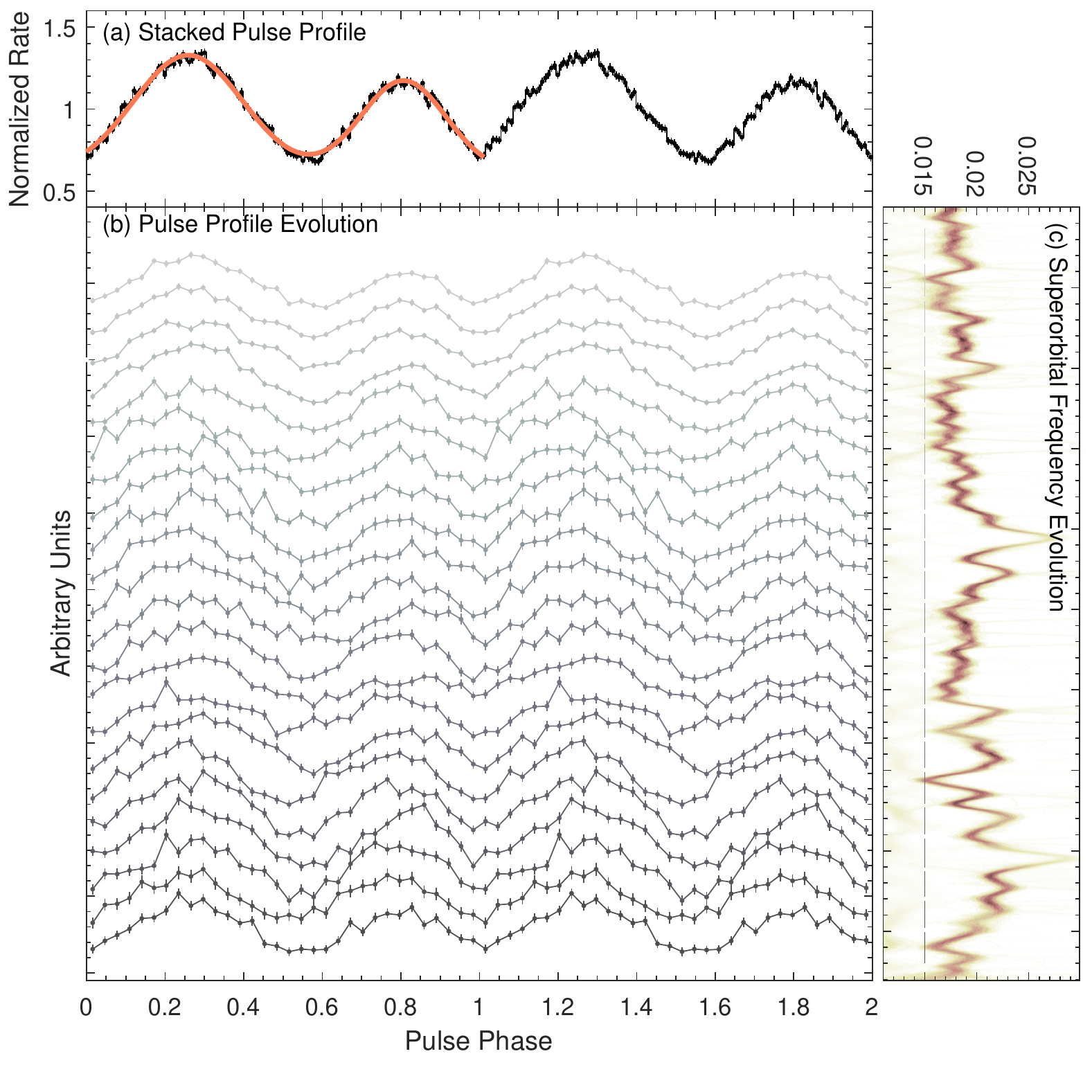}
 \caption{The 2--20 keV pulse profile evolution in the superorbital high state of SMC X-1. The stacked profile obtained from all 89 high states with MAXI is shown in panel (a). The bin size is 128 bins per cycle, and we plotted two cycles for visualization purposes.  The best-fit two-Gaussian model is plotted as the orange curve. The stacked profiles of every four consecutive superorbital high states are shown in panel (b) with 32 bins per cycle. Pulse profiles obtained from different epochs are denoted in different colours, where corresponding time intervals are shown as coloured straight lines in the stacked HHT spectrum of the superorbital modulation in panel (c). }
 \label{fig:pulse_profile_evolution}
\end{figure}

\subsection{Pulse Profile Evolution}
We investigate the long-term evolution of the 2--20 keV pulse profile using the MAXI data. We first accumulate all the photons collected in a total of 89 superorbital high states, fold them with their best local timing solution, and stack them together to obtain an averaged pulse profile, see Figure \ref{fig:pulse_profile_evolution} (a). The background level is estimated using photons collected within 1.5 -- 2 degrees from the position of SMC X-1, scaled to the same area in the sky and subtracted from the folded light curve. The pulse profile shows a typical double-peaked structure, consistent with previous works with pointed observations \citep[see, e.g.,][]{Neilsen2004, Hickox2005, BrumbackGB2022}. The primary peak lasts for $0.55$ cycles and another smaller peak lasts for $0.45$ cycles. The profile can be mathematically described by two Gaussian functions (Figure \ref{fig:pulse_profile_evolution} (a)).

To further trace the evolution of the pulse profile, we stack the profile every four superorbital cycles for visualization purposes and plot them in Figure \ref{fig:pulse_profile_evolution} (b). The overall structure does not change significantly although minor variability possibly exists. Such variability cannot be directly quantified by eye. Therefore, we parameterize the pulse profiles using three quantities; the pulsed fraction (PF), the phase separation and the ratio between two peaks, in order to follow the pulse profile evolution in more detail. We create pulse profiles from two consecutive superorbital cycles to keep enough photons in each profile. The PFs are calculated using the Fourier-based root-mean-square PF \citep[see definition and discussion in][]{DibKG2009, AnAH2015}. The separation and count rate ratio between the two peaks are obtained from the fitting result of two Gaussian functions. The uncertainties of all the parameters are obtained from $10^4$ times Monte Carlo simulations. 

The evolution of PF, peak separation, and count rate ratio are shown in Figure \ref{fig:superorbital_spin} (e),  (f), and (g), respectively. To further explore the evolutionary trends of parameters and eliminate short-term variability, we draw the EEMD band-pass filtered result and corresponding 1-$\sigma$ confidence intervals. The PF evolution is probably the most intriguing because it shows long-term variability, where the PF increases during both excursions 3 and 4. The other two parameters also show possible long-term variability although no clear connection to the superorbital or spin frequencies can be seen. We note that the pulse profile of SMC X-1 changes with both orbital {\it and} superorbital phases \citep[see, e.g.,][]{Naik2004, Neilsen2004, BrumbackGB2022}. Therefore, these results are only valid for long-term time-averaged behaviour. 

\section{Discussion}\label{discussion}

\subsection{The \nth{4} superorbital excursion}

Within the radiation-driven warping framework, SMC X-1 is expected to show a complex disc-warp configuration, likely a superposition of multiple warping modes and with a resultant warp shape that is time-variable, possibly in an aperiodic or quasi-periodic manner \citep{OgilvieD2001}. A time-variable warp shape would lead to time-variation in the phasing of features in the superorbital profile and a change in the instantaneous superorbital period. Both are strongly present in 1996--2021 interval analysed here (\autoref{fig:superorbital_shht}). 
Indeed, the superorbital modulation timescale as traced by the HHT indicates a number of ``mini-excursions'' (\autoref{fig:superorbital_shht} panels (a) and (b)). It is the apparent ramps in the phasing of the superorbital minima, followed by rapid, monotonic evolution to earlier superorbital phases, that suggest the superorbital excursions might be different in character from the mini-excursions.

Using changes in both the superorbital phasing and the instantaneous period as tracers, our analysis of the superorbital modulation shows a new excursion in 2020--2021 with a significantly shorter waiting time, longer duration, and a possibly shorter superorbital period in the pre-excursion epochs compared to previous excursions.
Under the warped-disc model, the superorbital modulation is caused by the variation in the absorption column and/or covering fraction of the central X-ray emission region. Then a change in the warp configuration would result in a change in the low-state flux, which has now been seen in all four superorbital excursions analyzed. Analysis of recent soft X-ray observations taken at a variety of superorbital phases as part of our program also shows corroborating evidence for this model (Karam et al. in prep). We note also that \citet{PradhanMP2020} find evidence from superorbital phase-resolved joint Suzaku/NuSTAR spectroscopic fits that indeed superorbital variation in partial covering fraction of the inner disc regions is a more likely mechanism than varying absorption column density. However, they go on to argue that the varying instantaneous accretion rate is a more likely mechanism for the superorbital modulation than any kind of varying obscuration. If indeed it is instantaneous $\dot{M}$~onto the accretor that varies with superorbital phase, then the lack of significant superorbital variation in the spin period (as opposed to the pulse profile) seen in \citet{PradhanMP2020} is puzzling.  

Dedicated soft X-ray observations in the superorbital low-state during and out-of-excursion epochs may be required to verify the change of the soft X-ray component, which would be contributed by the reprocessed X-rays from the inner accretion disc, in the low states during and out of excursions \citep{HickoxNK2004}.

The most intriguing feature of the \nth{4} excursion is the acceleration of the spin-up rate, which is not seen in the \nth{3} excursion but observed in the \nth{1} excursion.  A change in the accretion torque is expected to accompany a change in the warp inclination. The one-year time delay between the onset of the spin-up acceleration and the \nth{4} superorbital excursion implies a possible inside-out process between the change in accretion torque and the warp inclination. A similar time delay is possibly seen in the \nth{1} excursion \citep{DageCC2018}. 

\subsection{Variation in pulse profile}

The SMC X-1 pulse profile is thought to trace different components in the system. The {\it hard} pulse profile likely arises from a direct line of sight view of the accretor itself, while the {\it soft} pulsed component may come from reprocessing of hard X-ray emission by the inner disc \citep[][and references therein]{Neilsen2004, HickoxNK2004, BrumbackGB2022}. In a single 2-20 keV energy range, we do not have sensitivity to the hard and soft X-ray pulse profiles separately. All of the pulse profiles we report here were taken in the superorbital high state, corresponding to the most direct view of the accretor and maximum opening angle of the inner disc. Thus, our experiment presents 89 samples of the 2-20keV pulse profile with the inner disc in roughly the same orientation in three dimensions with respect to the line of sight.

Our pulse profile data provide coverage for the \nth{3} and \nth{4} excursions. There is evidence that the pulsed fraction increases for both superorbital excursions, which is generally consistent with a model in which the inner disc configuration indeed changes during the superorbital excursion (Figure \ref{fig:superorbital_spin}, panel e), or perhaps that the stream impact point(s) on the neutron star changes with superorbital cycle length. 

There is some room for improvement in the pulse profile analysis. For example, the timing solution is a simple first-order spin-up model within a few tens of days.  A time-variable timing noise between superorbital cycles would distort the observed pulse profile under such an oversimplified timing solution. 
Future observations that can derive a (semi-coherent) phase-connected timing solution in a shorter time scale would help clarify the evolution of the pulse profile. 

\subsection{Accretion and Disc variation}
Taken as a set, the four superorbital excursions do not suggest a simple, causative link between disc warp behaviour and accretion rate onto the neutron star that is suggested by the warped disc model \citep{Wijers1999, Still2004}.  While the \nth{1} and \nth{4} superorbital excursions are preceded by spin-up acceleration, the \nth{3} excursion does not show any pre-excursion spin-up trend: indeed, for the \nth{3} excursion the superorbital- and spin-modulations appear completely uncoupled (the \nth{2} excursion does not have sufficient spin-period coverage to make any inferences). This suggests that either the apparent pre-excursion spin-up increase of the neutron star in the \nth{1} and \nth{4} excursions is merely a coincidence, or there might be a threshold below which the two behaviours are not strongly coupled. The mechanism for such a threshold is not at present clear: from the limited sample of superorbital excursions, we note that the \nth{1} and \nth{4} excursions that show spin-up acceleration have a duration longer than the other two events. These may hint at various triggering mechanisms of the superorbital excursion. Another feature commonly seen in both the \nth{1} and \nth{4} excursions is the waiting time interval. The BATSE observation revealed a possible excursion event that ended at MJD 49100 (\nth{0} excursion) though the complete excursion was not clearly observed \citep[see Figure 2 in][]{ClarksonCC2003a}. If this is true and the duration of the \nth{0} excursion has a similar time scale of 400--800 days, the time interval between the \nth{0} and \nth{1} excursions would be roughly 1300--1700 days, similar to the time interval between the \nth{3} and \nth{4} excursions. We also note that a mini-excursion may occur before the 4th excursion, and the onset of the spin-up acceleration coincides with the end of this mini-excursion. Such mini excursions are probably seen in 2004, 2008, and 2012, but no corresponding change in the spin-up rate was observed in the 2012 event (see Figures \ref{fig:superorbital_shht} and \ref{fig:superorbital_spin}). Monitoring the spin-up rate and the superorbital period with all-sky monitoring programs would increase samples of superorbital excursions and help test their connection.

Finally, it was suggested that behaviour of SMC X-1's warp may be similar to that of Her X-1, which shows a connection between its pulse profile and superorbital modulation amplitude \citep{Still2004}. Her X-1's superorbital profile is double-peaked, with flux at the brighter peak correlated to the superorbital period \citep{Leahy2010}. Moreover, its spin frequency occasionally decreases, particularly during anomalous low states \citep{StaubertKP2009, StaubertKV2013}. These suggest that Her X-1's warp angle can sometimes be extremely high ($>90^{\circ}$), resulting in a negative accretion torque \citep{ParmarO1999}. Monitoring Her X-1's spin frequency change within one superorbital cycle reveals a sub-microsecond change with a 35 day period, suggesting a free precession of the axis of the neutron star\citep{KolesnikovSP2022}.  However, these behaviours are not observed in SMC X-1. The high-state flux remained stable during all excursions, implying a stable mass accretion rate. In addition, the increase in the low-state flux suggests a possible change in the warp or partial covering fraction, inferring a possible change in the spin frequency. Although both SMC X-1 and Her X-1 show variations in their soft pulse profile shape consistent with precession of a warped accretion disc \citep{Brumback2020, Brumback2021}, the binaries fundamentally differ in their mass transfer mechanisms. Her X-1's smaller A/F stellar companion transfers gas to the neutron star via Roche lobe overflow, whereas SMC X-1's B0 supergiant star likely also has stellar winds that complicate the mass transfer process \citep{ReynoldsQS1997, Webster1972}. Therefore, the superorbital modulation of SMC X-1 could be a combination of obscuring by warped accretion disc and other effects like the change in partial covering fraction of the inner disc \citep{PradhanMP2020, BrumbackGB2022}. Other models, like the precession of ring tube, the precession of funnel-shaped wind, or change in mass accretion rate, can possibly explain the superorbital modulation in SMC X-1 \citep{DauserMW2017, TsygankovMS2016, Inoue2019}. Future theoretical works are needed to test whether the observed phenomena, especially the connection between the spin and the superorbital modulations, can be observed in the model.

\section{Summary}\label{summary}
We perform a detailed analysis of the evolution of SMC X-1's superorbital modulation period and extend the time baseline to 26 years. We find a \nth{4} superorbital excursion, which has the longest duration and the shortest waiting time among all confirmed events. Moreover, we find a clear spin-up acceleration before the onset of the \nth{4} excursion, implying a possible inside-out process. This phenomenon is only seen in the \nth{1} and \nth{4} excursions but not in the \nth{3} event, suggesting possible different triggering mechanisms or some threshold that reveals the spin-superorbital connection. Monitoring the spin and superorbital behaviours of SMC X-1 and detailed timing and spectral observations in the following years are essential to reveal the underlying physics of the accretion in this system and provide links to other systems like PULXs. We encourage further theoretical studies to understand the conditions for producing observed connections between spin and superorbital modulations, and observations to test whether these behaviours can be seen in other systems.

\section*{Acknowledgements}
We thank the anonymous reviewer for valuable comments that improved this paper. This work made use of data provided by the ASM/RXTE teams at MIT and the RXTE SOF and GOF at NASA’s GSFC, the Swift BAT data provided by the hard X-ray transient monitor \citep{KrimmHC2013}, and the MAXI data provided by RIKEN, JAXA, and the MAXI team. CPH acknowledges support from the National Science and Technology Council in Taiwan through grant MOST 109-2112-M-018-009-MY3.  KCD acknowledges fellowship funding from Fonds de Recherche du Qu\'ebec $-$ Nature et Technologies, Bourses de recherche postdoctorale B3X no. 319864. DH acknowledges funding from the Natural Sciences and Engineering Research Council of Canada (NSERC) and the Canada Research Chairs (CRC) program.

\section*{DATA AVAILABILITY}

The \rxte\ ASM light curve can be downloaded from the ASM Source Catalog (\url{https://heasarc.gsfc.nasa.gov/docs/xte/ASM/sources.html}). The \swift\ BAT light curve \citep{KrimmHC2013} is  publicly available from the \swift/BAT Hard X-ray Transient Monitor program webpage (\url{https://swift.gsfc.nasa.gov/results/transients/}).
The MAXI GSC data can be downloaded and processed using the MAXI software package, which has been implemented within HEASoft (\url{https://heasarc.gsfc.nasa.gov/docs/software/heasoft/}) since version 6.25. The spin frequency measurements of this work are archived on this webpage maintained by the MAXI team at RIKEN (\url{http://maxi.riken.jp/pulsar/smcx1/}).

\bibliographystyle{mnras}

\begin{thebibliography}{}
\makeatletter
\relax
\def\mn@urlcharsother{\let\do\@makeother \do\$\do\&\do\#\do\^\do\_\do\%\do\~}
\def\mn@doi{\begingroup\mn@urlcharsother \@ifnextchar [ {\mn@doi@}
  {\mn@doi@[]}}
\def\mn@doi@[#1]#2{\def\@tempa{#1}\ifx\@tempa\@empty \href
  {http://dx.doi.org/#2} {doi:#2}\else \href {http://dx.doi.org/#2} {#1}\fi
  \endgroup}
\def\mn@eprint#1#2{\mn@eprint@#1:#2::\@nil}
\def\mn@eprint@arXiv#1{\href {http://arxiv.org/abs/#1} {{\tt arXiv:#1}}}
\def\mn@eprint@dblp#1{\href {http://dblp.uni-trier.de/rec/bibtex/#1.xml}
  {dblp:#1}}
\def\mn@eprint@#1:#2:#3:#4\@nil{\def\@tempa {#1}\def\@tempb {#2}\def\@tempc
  {#3}\ifx \@tempc \@empty \let \@tempc \@tempb \let \@tempb \@tempa \fi \ifx
  \@tempb \@empty \def\@tempb {arXiv}\fi \@ifundefined
  {mn@eprint@\@tempb}{\@tempb:\@tempc}{\expandafter \expandafter \csname
  mn@eprint@\@tempb\endcsname \expandafter{\@tempc}}}

\bibitem[\protect\citeauthoryear{{An} et~al.,}{{An} et~al.}{2015}]{AnAH2015}
{An} H.,  et~al., 2015, \mn@doi [\apj] {10.1088/0004-637X/807/1/93}, \href
  {http://ui.adsabs.harvard.edu/abs/2015ApJ...807...93A} {807, 93}

\bibitem[\protect\citeauthoryear{{Bachetti} et~al.,}{{Bachetti}
  et~al.}{2014}]{BachettiHW2014}
{Bachetti} M.,  et~al., 2014, \mn@doi [\nat] {10.1038/nature13791}, \href
  {http://ui.adsabs.harvard.edu/abs/2014Natur.514..202B} {514, 202}

\bibitem[\protect\citeauthoryear{{Barthelmy} et~al.,}{{Barthelmy}
  et~al.}{2005}]{BarthelmyBC2005}
{Barthelmy} S.~D.,  et~al., 2005, \mn@doi [\ssr] {10.1007/s11214-005-5096-3},
  \href {http://ui.adsabs.harvard.edu/abs/2005SSRv..120..143B} {120, 143}

\bibitem[\protect\citeauthoryear{{Brightman} et~al.,}{{Brightman}
  et~al.}{2020}]{BrightmanEF2020}
{Brightman} M.,  et~al., 2020, \mn@doi [\apj] {10.3847/1538-4357/ab7e2a}, \href
  {https://ui.adsabs.harvard.edu/abs/2020ApJ...895..127B} {895, 127}

\bibitem[\protect\citeauthoryear{{Brumback}, {Hickox}, {F{\"u}rst},
  {Pottschmidt}, {Tomsick}  \& {Wilms}}{{Brumback} et~al.}{2020}]{Brumback2020}
{Brumback} M.~C.,  {Hickox} R.~C.,  {F{\"u}rst} F.~S.,  {Pottschmidt} K.,
  {Tomsick} J.~A.,   {Wilms} J.,  2020, \mn@doi [\apj]
  {10.3847/1538-4357/ab5b0410.48550/arXiv.1909.10559}, \href
  {https://ui.adsabs.harvard.edu/abs/2020ApJ...888..125B} {888, 125}

\bibitem[\protect\citeauthoryear{{Brumback}, {Hickox}, {F{\"u}rst},
  {Pottschmidt}, {Tomsick}, {Wilms}, {Staubert}  \& {Vrtilek}}{{Brumback}
  et~al.}{2021}]{Brumback2021}
{Brumback} M.~C.,  {Hickox} R.~C.,  {F{\"u}rst} F.~S.,  {Pottschmidt} K.,
  {Tomsick} J.~A.,  {Wilms} J.,  {Staubert} R.,   {Vrtilek} S.,  2021, \mn@doi
  [\apj] {10.3847/1538-4357/abe12210.48550/arXiv.2102.05097}, \href
  {https://ui.adsabs.harvard.edu/abs/2021ApJ...909..186B} {909, 186}

\bibitem[\protect\citeauthoryear{{Brumback} et~al.,}{{Brumback}
  et~al.}{2022}]{BrumbackGB2022}
{Brumback} M.~C.,  et~al., 2022, \mn@doi [\apj] {10.3847/1538-4357/ac4d24},
  \href {https://ui.adsabs.harvard.edu/abs/2022ApJ...926..187B} {926, 187}

\bibitem[\protect\citeauthoryear{{Buccheri} et~al.,}{{Buccheri}
  et~al.}{1983}]{Buccheri1983}
{Buccheri} R.,  et~al., 1983, \aap, \href
  {http://ui.adsabs.harvard.edu/abs/1983A%26A...128..245B} {128, 245}

\bibitem[\protect\citeauthoryear{{Charles}, {Clarkson}, {Cornelisse}  \&
  {Shih}}{{Charles} et~al.}{2008}]{Charles2008}
{Charles} P.,  {Clarkson} W.,  {Cornelisse} R.,   {Shih} C.,  2008, \mn@doi
  [New Astronomy Reviews] {10.1016/j.newar.2008.03.025}, \href
  {http://ui.adsabs.harvard.edu/abs/2008NewAR..51..768C} {51, 768}

\bibitem[\protect\citeauthoryear{{Clarkson}, {Charles}, {Coe}, {Laycock},
  {Tout}  \& {Wilson}}{{Clarkson} et~al.}{2003a}]{ClarksonCC2003a}
{Clarkson} W.~I.,  {Charles} P.~A.,  {Coe} M.~J.,  {Laycock} S.,  {Tout} M.~D.,
    {Wilson} C.~A.,  2003a, \mn@doi [\mnras]
  {10.1046/j.1365-8711.2003.06176.x}, \href
  {http://ui.adsabs.harvard.edu/abs/2003MNRAS.339..447C} {339, 447}

\bibitem[\protect\citeauthoryear{{Clarkson}, {Charles}, {Coe}  \&
  {Laycock}}{{Clarkson} et~al.}{2003b}]{ClarksonCC2003b}
{Clarkson} W.~I.,  {Charles} P.~A.,  {Coe} M.~J.,   {Laycock} S.,  2003b,
  \mn@doi [\mnras] {10.1046/j.1365-8711.2003.06761.x}, \href
  {http://ui.adsabs.harvard.edu/abs/2003MNRAS.343.1213C} {343, 1213}

\bibitem[\protect\citeauthoryear{{Dage}, {Clarkson}, {Charles}, {Laycock}  \&
  {Shih}}{{Dage} et~al.}{2019}]{DageCC2018}
{Dage} K.~C.,  {Clarkson} W.~I.,  {Charles} P.~A.,  {Laycock} S.~G.~T.,
  {Shih} I.-C.,  2019, \mn@doi [\mnras] {10.1093/mnras/sty2572}, \href
  {http://ui.adsabs.harvard.edu/abs/2019MNRAS.482..337D} {482, 337}

\bibitem[\protect\citeauthoryear{{Dage} et~al.,}{{Dage}
  et~al.}{2022}]{DageBN2022}
{Dage} K.~C.,  et~al., 2022, \mn@doi [\mnras] {10.1093/mnras/stac1674}, \href
  {https://ui.adsabs.harvard.edu/abs/2022MNRAS.514.5457D} {514, 5457}

\bibitem[\protect\citeauthoryear{{Dauser}, {Middleton}  \& {Wilms}}{{Dauser}
  et~al.}{2017}]{DauserMW2017}
{Dauser} T.,  {Middleton} M.,   {Wilms} J.,  2017, \mn@doi [\mnras]
  {10.1093/mnras/stw3304}, \href
  {http://ui.adsabs.harvard.edu/abs/2017MNRAS.466.2236D} {466, 2236}

\bibitem[\protect\citeauthoryear{{Dib}, {Kaspi}  \& {Gavriil}}{{Dib}
  et~al.}{2009}]{DibKG2009}
{Dib} R.,  {Kaspi} V.~M.,   {Gavriil} F.~P.,  2009, \mn@doi [\apj]
  {10.1088/0004-637X/702/1/614}, \href
  {http://ui.adsabs.harvard.edu/abs/2009ApJ...702..614D} {702, 614}

\bibitem[\protect\citeauthoryear{{Falanga}, {Bozzo}, {Lutovinov},
  {Bonnet-Bidaud}, {Fetisova}  \& {Puls}}{{Falanga} et~al.}{2015}]{Falanga2015}
{Falanga} M.,  {Bozzo} E.,  {Lutovinov} A.,  {Bonnet-Bidaud} J.~M.,  {Fetisova}
  Y.,   {Puls} J.,  2015, \mn@doi [\aap] {10.1051/0004-6361/201425191}, \href
  {http://ui.adsabs.harvard.edu/abs/2015A%26A...577A.130F} {577, A130}

\bibitem[\protect\citeauthoryear{{Foulkes}, {Haswell}  \& {Murray}}{{Foulkes}
  et~al.}{2010}]{FoulkesHM2010}
{Foulkes} S.~B.,  {Haswell} C.~A.,   {Murray} J.~R.,  2010, \mn@doi [\mnras]
  {10.1111/j.1365-2966.2009.15721.x}, \href
  {https://ui.adsabs.harvard.edu/abs/2010MNRAS.401.1275F} {401, 1275}

\bibitem[\protect\citeauthoryear{{Gruber} \& {Rothschild}}{{Gruber} \&
  {Rothschild}}{1984}]{Gruber1984}
{Gruber} D.~E.,  {Rothschild} R.~E.,  1984, \mn@doi [\apj] {10.1086/162338},
  \href {http://ui.adsabs.harvard.edu/abs/1984ApJ...283..546G} {283, 546}

\bibitem[\protect\citeauthoryear{{Hickox} \& {Vrtilek}}{{Hickox} \&
  {Vrtilek}}{2005}]{Hickox2005}
{Hickox} R.~C.,  {Vrtilek} S.~D.,  2005, \mn@doi [\apj] {10.1086/491596}, \href
  {http://ui.adsabs.harvard.edu/abs/2005ApJ...633.1064H} {633, 1064}

\bibitem[\protect\citeauthoryear{{Hickox}, {Narayan}  \& {Kallman}}{{Hickox}
  et~al.}{2004}]{HickoxNK2004}
{Hickox} R.~C.,  {Narayan} R.,   {Kallman} T.~R.,  2004, \mn@doi [\apj]
  {10.1086/423928}, \href
  {https://ui.adsabs.harvard.edu/abs/2004ApJ...614..881H} {614, 881}

\bibitem[\protect\citeauthoryear{{Hu}, {Chou}, {Wu}, {Yang}  \& {Su}}{{Hu}
  et~al.}{2011}]{Hu2011}
{Hu} C.-P.,  {Chou} Y.,  {Wu} M.-C.,  {Yang} T.-C.,   {Su} Y.-H.,  2011,
  \mn@doi [\apj] {10.1088/0004-637X/740/2/67}, \href
  {http://ui.adsabs.harvard.edu/abs/2011ApJ...740...67H} {740, 67}

\bibitem[\protect\citeauthoryear{{Hu}, {Li}, {Kong}, {Ng}  \& {Chun-Che
  Lin}}{{Hu} et~al.}{2017}]{HuLK2017}
{Hu} C.-P.,  {Li} K.~L.,  {Kong} A.~K.~H.,  {Ng} C.-Y.,   {Chun-Che Lin} L.,
  2017, \mn@doi [\apjl] {10.3847/2041-8213/835/1/L9}, \href
  {http://ui.adsabs.harvard.edu/abs/2017ApJ...835L...9H} {835, L9}

\bibitem[\protect\citeauthoryear{{Hu}, {Mihara}, {Sugizaki}, {Ueda}  \&
  {Enoto}}{{Hu} et~al.}{2019}]{HuMS2019}
{Hu} C.-P.,  {Mihara} T.,  {Sugizaki} M.,  {Ueda} Y.,   {Enoto} T.,  2019,
  \mn@doi [\apj] {10.3847/1538-4357/ab48e4}, \href
  {https://ui.adsabs.harvard.edu/abs/2019ApJ...885..123H} {885, 123}

\bibitem[\protect\citeauthoryear{{Hu}, {Lin}, {Pan}, {Li}, {Yen}, {Kong}  \&
  {Hui}}{{Hu} et~al.}{2022}]{HuLP2022}
{Hu} C.-P.,  {Lin} L. C.-C.,  {Pan} K.-C.,  {Li} K.-L.,  {Yen} C.-C.,  {Kong}
  A. K.~H.,   {Hui} C.~Y.,  2022, \mn@doi [\apj] {10.3847/1538-4357/ac8165},
  \href {https://ui.adsabs.harvard.edu/abs/2022ApJ...935..127H} {935, 127}

\bibitem[\protect\citeauthoryear{{Huang} et~al.,}{{Huang}
  et~al.}{1998}]{Huang1998}
{Huang} N.~E.,  et~al., 1998, \mn@doi [Royal Society of London Proceedings
  Series A] {10.1098/rspa.1998.0193}, \href
  {http://ui.adsabs.harvard.edu/abs/1998RSPSA.454..903E} {454, 903}

\bibitem[\protect\citeauthoryear{\.{I}nam, Baykal  \& Beklen}{\.{I}nam
  et~al.}{2010}]{Inam2010}
\.{I}nam S.~c.,  Baykal A.,   Beklen E.,  2010, \mn@doi [\mnras]
  {10.1111/j.1365-2966.2009.16121.x}, \href
  {http://ui.adsabs.harvard.edu/abs/2010MNRAS.403..378I} {403, 378}

\bibitem[\protect\citeauthoryear{{Inoue}}{{Inoue}}{2019}]{Inoue2019}
{Inoue} H.,  2019, \mn@doi [\pasj] {10.1093/pasj/psy152}, \href
  {http://ui.adsabs.harvard.edu/abs/2019PASJ..tmp...31I} {}

\bibitem[\protect\citeauthoryear{Kahaner, Moler  \& Nash}{Kahaner
  et~al.}{1989}]{Kahaner1989}
Kahaner D.,  Moler C.,   Nash S.,  1989, Numerical methods and software

\bibitem[\protect\citeauthoryear{{King} \& {Lasota}}{{King} \&
  {Lasota}}{2019}]{KingL2019}
{King} A.,  {Lasota} J.-P.,  2019, \mn@doi [\mnras] {10.1093/mnras/stz720},
  \href {https://ui.adsabs.harvard.edu/abs/2019MNRAS.485.3588K} {485, 3588}

\bibitem[\protect\citeauthoryear{{Kolesnikov}, {Shakura}  \&
  {Postnov}}{{Kolesnikov} et~al.}{2022}]{KolesnikovSP2022}
{Kolesnikov} D.,  {Shakura} N.,   {Postnov} K.,  2022, \mn@doi [\mnras]
  {10.1093/mnras/stac1107}, \href
  {https://ui.adsabs.harvard.edu/abs/2022MNRAS.513.3359K} {513, 3359}

\bibitem[\protect\citeauthoryear{{Kong}, {Hu}, {Lin}, {Li}, {Jin}, {Liu}  \&
  {Yen}}{{Kong} et~al.}{2016}]{KongHL2016}
{Kong} A.~K.~H.,  {Hu} C.-P.,  {Lin} L.~C.-C.,  {Li} K.~L.,  {Jin} R.,  {Liu}
  C.~Y.,   {Yen} D.~C.-C.,  2016, \mn@doi [\mnras] {10.1093/mnras/stw1558},
  \href {http://ui.adsabs.harvard.edu/abs/2016MNRAS.461.4395K} {461, 4395}

\bibitem[\protect\citeauthoryear{{Krimm} et~al.,}{{Krimm}
  et~al.}{2013}]{KrimmHC2013}
{Krimm} H.~A.,  et~al., 2013, \mn@doi [\apjs] {10.1088/0067-0049/209/1/14},
  \href {http://ui.adsabs.harvard.edu/abs/2013ApJS..209...14K} {209, 14}

\bibitem[\protect\citeauthoryear{{Leahy} \& {Igna}}{{Leahy} \&
  {Igna}}{2010}]{Leahy2010}
{Leahy} D.~A.,  {Igna} C.~D.,  2010, \mn@doi [\apj]
  {10.1088/0004-637X/713/1/318}, \href
  {http://ui.adsabs.harvard.edu/abs/2010ApJ...713..318L} {713, 318}

\bibitem[\protect\citeauthoryear{{Levine}, {Bradt}, {Cui}, {Jernigan},
  {Morgan}, {Remillard}, {Shirey}  \& {Smith}}{{Levine}
  et~al.}{1996}]{LevineBC1996}
{Levine} A.~M.,  {Bradt} H.,  {Cui} W.,  {Jernigan} J.~G.,  {Morgan} E.~H.,
  {Remillard} R.,  {Shirey} R.~E.,   {Smith} D.~A.,  1996, \mn@doi [\apjl]
  {10.1086/310260}, \href
  {http://ui.adsabs.harvard.edu/abs/1996ApJ...469L..33L} {469, L33}

\bibitem[\protect\citeauthoryear{{Lucke}, {Yentis}, {Friedman}, {Fritz}  \&
  {Shulman}}{{Lucke} et~al.}{1976}]{Lucke1976}
{Lucke} R.,  {Yentis} D.,  {Friedman} H.,  {Fritz} G.,   {Shulman} S.,  1976,
  \mn@doi [\apjl] {10.1086/182125}, \href
  {http://ui.adsabs.harvard.edu/abs/1976ApJ...206L..25L} {206, L25}

\bibitem[\protect\citeauthoryear{{Matsuoka} et~al.,}{{Matsuoka}
  et~al.}{2009}]{MatsuokaKU2009}
{Matsuoka} M.,  et~al., 2009, \mn@doi [\pasj] {10.1093/pasj/61.5.999}, \href
  {http://ads.nao.ac.jp/abs/2009PASJ...61..999M} {61, 999}

\bibitem[\protect\citeauthoryear{{Mihara} et~al.,}{{Mihara}
  et~al.}{2011}]{MiharaNS2011}
{Mihara} T.,  et~al., 2011, \mn@doi [\pasj] {10.1093/pasj/63.sp3.S623}, \href
  {http://ui.adsabs.harvard.edu/abs/2011PASJ...63S.623M} {63, S623}

\bibitem[\protect\citeauthoryear{{Mihara}, {Tsunemi}  \& {Negoro}}{{Mihara}
  et~al.}{2022}]{MihTN2022}
{Mihara} T.,  {Tsunemi} H.,   {Negoro} H.,  2022, arXiv e-prints, \href
  {https://ui.adsabs.harvard.edu/abs/2022arXiv220601505M} {p. arXiv:2206.01505}

\bibitem[\protect\citeauthoryear{{Naik} \& {Paul}}{{Naik} \&
  {Paul}}{2004}]{Naik2004}
{Naik} S.,  {Paul} B.,  2004, \mn@doi [\aap] {10.1051/0004-6361:20040048},
  \href {http://ui.adsabs.harvard.edu/abs/2004A%26A...418..655N} {418, 655}

\bibitem[\protect\citeauthoryear{{Neilsen}, {Hickox}  \& {Vrtilek}}{{Neilsen}
  et~al.}{2004}]{Neilsen2004}
{Neilsen} J.,  {Hickox} R.~C.,   {Vrtilek} S.~D.,  2004, \mn@doi [\apjl]
  {10.1086/426890}, \href
  {http://ui.adsabs.harvard.edu/abs/2004ApJ...616L.135N} {616, L135}

\bibitem[\protect\citeauthoryear{{Ogilvie} \& {Dubus}}{{Ogilvie} \&
  {Dubus}}{2001}]{OgilvieD2001}
{Ogilvie} G.~I.,  {Dubus} G.,  2001, \mn@doi [\mnras]
  {10.1046/j.1365-8711.2001.04011.x}, \href
  {http://ui.adsabs.harvard.edu/abs/2001MNRAS.320..485O} {320, 485}

\bibitem[\protect\citeauthoryear{{Parmar}, {Oosterbroek}, {dal Fiume},
  {Orlandini}, {Santangelo}, {Segreto}  \& {del Sordo}}{{Parmar}
  et~al.}{1999}]{ParmarO1999}
{Parmar} A.~N.,  {Oosterbroek} T.,  {dal Fiume} D.,  {Orlandini} M.,
  {Santangelo} A.,  {Segreto} A.,   {del Sordo} S.,  1999, \aap, \href
  {http://ui.adsabs.harvard.edu/abs/1999A%26A...350L...5P} {350, L5}

\bibitem[\protect\citeauthoryear{{Pike} et~al.,}{{Pike}
  et~al.}{2019}]{PikeHB2019}
{Pike} S.~N.,  et~al., 2019, \mn@doi [\apj] {10.3847/1538-4357/ab0f2b}, \href
  {https://ui.adsabs.harvard.edu/abs/2019ApJ...875..144P} {875, 144}

\bibitem[\protect\citeauthoryear{{Pradhan}, {Maitra}  \& {Paul}}{{Pradhan}
  et~al.}{2020}]{PradhanMP2020}
{Pradhan} P.,  {Maitra} C.,   {Paul} B.,  2020, \mn@doi [\apj]
  {10.3847/1538-4357/ab8224}, \href
  {https://ui.adsabs.harvard.edu/abs/2020ApJ...895...10P} {895, 10}

\bibitem[\protect\citeauthoryear{{Pringle}}{{Pringle}}{1996}]{Pringle1996}
{Pringle} J.~E.,  1996, \mn@doi [\mnras] {10.1093/mnras/281.1.357}, \href
  {http://ui.adsabs.harvard.edu/abs/1996MNRAS.281..357P} {281, 357}

\bibitem[\protect\citeauthoryear{{Reynolds}, {Hilditch}, {Bell}  \&
  {Hill}}{{Reynolds} et~al.}{1993}]{Reynolds1993}
{Reynolds} A.~P.,  {Hilditch} R.~W.,  {Bell} S.~A.,   {Hill} G.,  1993, \mnras,
  \href {http://ui.adsabs.harvard.edu/abs/1993MNRAS.261..337R} {261, 337}

\bibitem[\protect\citeauthoryear{{Reynolds}, {Quaintrell}, {Still}, {Roche},
  {Chakrabarty}  \& {Levine}}{{Reynolds} et~al.}{1997}]{ReynoldsQS1997}
{Reynolds} A.~P.,  {Quaintrell} H.,  {Still} M.~D.,  {Roche} P.,  {Chakrabarty}
  D.,   {Levine} S.~E.,  1997, \mn@doi [\mnras] {10.1093/mnras/288.1.43}, \href
  {http://ui.adsabs.harvard.edu/abs/1997MNRAS.288...43R} {288, 43}

\bibitem[\protect\citeauthoryear{{Staubert}, {Klochkov}, {Postnov}, {Shakura},
  {Wilms}  \& {Rothschild}}{{Staubert} et~al.}{2009}]{StaubertKP2009}
{Staubert} R.,  {Klochkov} D.,  {Postnov} K.,  {Shakura} N.,  {Wilms} J.,
  {Rothschild} R.~E.,  2009, \mn@doi [\aap] {10.1051/0004-6361:200810743},
  \href {http://ui.adsabs.harvard.edu/abs/2009A%26A...494.1025S} {494, 1025}

\bibitem[\protect\citeauthoryear{{Staubert}, {Klochkov}, {Vasco}, {Postnov},
  {Shakura}, {Wilms}  \& {Rothschild}}{{Staubert}
  et~al.}{2013}]{StaubertKV2013}
{Staubert} R.,  {Klochkov} D.,  {Vasco} D.,  {Postnov} K.,  {Shakura} N.,
  {Wilms} J.,   {Rothschild} R.~E.,  2013, \mn@doi [\aap]
  {10.1051/0004-6361/201220316}, \href
  {http://ui.adsabs.harvard.edu/abs/2013A%26A...550A.110S} {550, A110}

\bibitem[\protect\citeauthoryear{{Still} \& {Boyd}}{{Still} \&
  {Boyd}}{2004}]{Still2004}
{Still} M.,  {Boyd} P.,  2004, \mn@doi [\apjl] {10.1086/421349}, \href
  {http://ui.adsabs.harvard.edu/abs/2004ApJ...606L.135S} {606, L135}

\bibitem[\protect\citeauthoryear{{Sugizaki} et~al.,}{{Sugizaki}
  et~al.}{2011}]{SugizakiMS2011}
{Sugizaki} M.,  et~al., 2011, \mn@doi [\pasj] {10.1093/pasj/63.sp3.S635}, \href
  {http://ui.adsabs.harvard.edu/abs/2011PASJ...63S.635S} {63, S635}

\bibitem[\protect\citeauthoryear{{Thomas}, {Charles}, {Buckley}, {Kotze},
  {Lasota}, {Potter}, {Steiner}  \& {Paice}}{{Thomas}
  et~al.}{2022}]{ThomasCB2022}
{Thomas} J.~K.,  {Charles} P.~A.,  {Buckley} D. A.~H.,  {Kotze} M.~M.,
  {Lasota} J.-P.,  {Potter} S.~B.,  {Steiner} J.~F.,   {Paice} J.~A.,  2022,
  \mn@doi [\mnras] {10.1093/mnras/stab3033}, \href
  {https://ui.adsabs.harvard.edu/abs/2022MNRAS.509.1062T} {509, 1062}

\bibitem[\protect\citeauthoryear{{Trowbridge}, {Nowak}  \&
  {Wilms}}{{Trowbridge} et~al.}{2007}]{Trowbridge2007}
{Trowbridge} S.,  {Nowak} M.~A.,   {Wilms} J.,  2007, \mn@doi [\apj]
  {10.1086/522075}, \href
  {http://ui.adsabs.harvard.edu/abs/2007ApJ...670..624T} {670, 624}

\bibitem[\protect\citeauthoryear{{Tsygankov}, {Mushtukov}, {Suleimanov}  \&
  {Poutanen}}{{Tsygankov} et~al.}{2016}]{TsygankovMS2016}
{Tsygankov} S.~S.,  {Mushtukov} A.~A.,  {Suleimanov} V.~F.,   {Poutanen} J.,
  2016, \mn@doi [\mnras] {10.1093/mnras/stw046}, \href
  {https://ui.adsabs.harvard.edu/abs/2016MNRAS.457.1101T} {457, 1101}

\bibitem[\protect\citeauthoryear{{Vasilopoulos}, {Koliopanos}, {Haberl},
  {Treiber}, {Brightman}, {Earnshaw}  \& {G{\'u}rpide}}{{Vasilopoulos}
  et~al.}{2021}]{VasilopoulosHT2021}
{Vasilopoulos} G.,  {Koliopanos} F.,  {Haberl} F.,  {Treiber} H.,  {Brightman}
  M.,  {Earnshaw} H.~P.,   {G{\'u}rpide} A.,  2021, \mn@doi [\apj]
  {10.3847/1538-4357/abda49}, \href
  {https://ui.adsabs.harvard.edu/abs/2021ApJ...909...50V} {909, 50}

\bibitem[\protect\citeauthoryear{{Walton} et~al.,}{{Walton}
  et~al.}{2016}]{WaltonFB2016}
{Walton} D.~J.,  et~al., 2016, \mn@doi [\apjl] {10.3847/2041-8205/827/1/L13},
  \href {http://ui.adsabs.harvard.edu/abs/2016ApJ...827L..13W} {827, L13}

\bibitem[\protect\citeauthoryear{{Wang}, {Yeh}, {Young}, {Hu}  \& {Lo}}{{Wang}
  et~al.}{2014}]{WangYY2014}
{Wang} Y.-H.,  {Yeh} C.-H.,  {Young} H.-W.~V.,  {Hu} K.,   {Lo} M.-T.,  2014,
  \mn@doi [Physica A Statistical Mechanics and its Applications]
  {10.1016/j.physa.2014.01.020}, \href
  {http://ui.adsabs.harvard.edu/abs/2014PhyA..400..159W} {400, 159}

\bibitem[\protect\citeauthoryear{{Webster}, {Martin}, {Feast}  \&
  {Andrews}}{{Webster} et~al.}{1972}]{Webster1972}
{Webster} B.~L.,  {Martin} W.~L.,  {Feast} M.~W.,   {Andrews} P.~J.,  1972,
  \mn@doi [Nature Physical Science] {10.1038/physci240183b0}, \href
  {https://ui.adsabs.harvard.edu/abs/1972NPhS..240R.183W} {240, 183}

\bibitem[\protect\citeauthoryear{{Wijers} \& {Pringle}}{{Wijers} \&
  {Pringle}}{1999}]{Wijers1999}
{Wijers} R.~A.~M.~J.,  {Pringle} J.~E.,  1999, \mn@doi [\mnras]
  {10.1046/j.1365-8711.1999.02720.x}, \href
  {http://ui.adsabs.harvard.edu/abs/1999MNRAS.308..207W} {308, 207}

\bibitem[\protect\citeauthoryear{{Wojdowski}, {Clark}, {Levine}, {Woo}  \&
  {Zhang}}{{Wojdowski} et~al.}{1998}]{Wojdowski1998}
{Wojdowski} P.,  {Clark} G.~W.,  {Levine} A.~M.,  {Woo} J.~W.,   {Zhang} S.~N.,
   1998, \mn@doi [\apj] {10.1086/305893}, \href
  {http://ui.adsabs.harvard.edu/abs/1998ApJ...502..253W} {502, 253}

\bibitem[\protect\citeauthoryear{Wu \& Huang}{Wu \& Huang}{2009}]{Wu2009}
Wu Z.,  Huang N.~E.,  2009, \mn@doi [AADA] {10.1142/S1793536909000047}, 1, 1

\bibitem[\protect\citeauthoryear{{Yeh}, {Shieh}  \& {Huang}}{{Yeh}
  et~al.}{2010}]{YehSH2010}
{Yeh} J.-R.,  {Shieh} J.-S.,   {Huang} N.~E.,  2010, \mn@doi [Advances in
  Adaptive Data Analysis] {10.1142/S1793536910000422}, 02, 135

\bibitem[\protect\citeauthoryear{{van der Meer}, {Kaper}, {van Kerkwijk},
  {Heemskerk}  \& {van den Heuvel}}{{van der Meer}
  et~al.}{2007}]{vanderMeer2007}
{van der Meer} A.,  {Kaper} L.,  {van Kerkwijk} M.~H.,  {Heemskerk} M.~H.~M.,
  {van den Heuvel} E.~P.~J.,  2007, \mn@doi [\aap]
  {10.1051/0004-6361:20066025}, \href
  {http://ui.adsabs.harvard.edu/abs/2007A%26A...473..523V} {473, 523}

\makeatother
\end{thebibliography}
\input{output.bbl}


\bsp	
\label{lastpage}
\end{document}